\documentclass[aps,pra,twocolumn,amsmath,amsmath,amssymb,amsfonts,10pt,superscriptaddress]{revtex4-1}
\usepackage{fullpage}
\usepackage[utf8]{inputenc}
\usepackage{amsmath}
\usepackage{mathtools}
\usepackage{sfmath}
\usepackage{color}
\usepackage{amsfonts}
\usepackage{amssymb}
\usepackage{bbold}
\usepackage{graphicx}
\usepackage{caption}
\usepackage{subcaption}
\definecolor{darkgreen}{rgb}{0.0, 0.5, 0.0}

\newcommand\scalemath[2]{\scalebox{#1}{\mbox{\ensuremath{\displaystyle #2}}}}
\allowdisplaybreaks

\begin{document}
\title{Quantum estimation of coupling strengths in driven-dissipative optomechanics}
\author{Kamila Sala}
\affiliation{Centre for the Mathematics and Theoretical Physics of Quantum Non-Equilibrium Systems, University of Nottingham, Nottingham, NG7 2RD, United Kingdom}
\affiliation{School of Mathematical Sciences, University of Nottingham, Nottingham, NG7 2RD, United Kingdom}
\author{Anton Doicin}
\affiliation{School of Mathematical Sciences, University of Nottingham, Nottingham, NG7 2RD, United Kingdom}
\author{Andrew D. Armour}
\affiliation{Centre for the Mathematics and Theoretical Physics of Quantum Non-Equilibrium Systems, University of Nottingham, Nottingham, NG7 2RD, United Kingdom}
\affiliation{School of Physics and Astronomy, University of Nottingham, Nottingham, NG7 2RD, United Kingdom}

\author{Tommaso Tufarelli}
\affiliation{Centre for the Mathematics and Theoretical Physics of Quantum Non-Equilibrium Systems, University of Nottingham, Nottingham, NG7 2RD, United Kingdom}
\affiliation{School of Mathematical Sciences, University of Nottingham, Nottingham, NG7 2RD, United Kingdom}
\begin{abstract}
We exploit local quantum estimation theory to investigate the measurement of linear and quadratic coupling strengths in a driven-dissipative optomechanical system. For experimentally realistic values of the model parameters, we 
find that the linear coupling strength is considerably easier to estimate than the quadratic one. Our analysis also reveals that the majority of information about these parameters is encoded in the reduced state of the mechanical element, and that the best estimation strategy for both coupling parameters is well approximated by a direct measurement of the mechanical position quadrature.
Interestingly, we also show that temperature does not always have a detrimental effect on the estimation precision, and that the effects of temperature are more pronounced in the case of the quadratic coupling parameter.    
\end{abstract}
\maketitle     

\section{Introduction} 

Quantum optomechanics focuses on the interaction between the electromagnetic radiation and motional degrees of freedom of mechanical oscillators\,\cite{Aspelmeyer,optobook,backaction}.  The simplest optomechanical system consists of a single cavity mode interacting with a single mechanical mode and is realised, for example, in an optical cavity with a movable mirror. In this case the mechanism responsible for the interaction is radiation pressure, which entails momentum exchange between light and matter. The presence of a cavity boosts the otherwise weak radiation pressure force, enhancing the light-matter interaction. The quantum effects of radiation pressure forces and the associated limits they set on the precision of mirror-displacement measurements are of great importance for many applications including gravitational wave detectors, scanning probe microscopy and force sensing\,\cite{classical,backaction,optobook,Aspelmeyer}. 

Although the radiation pressure interaction is intrinsically non-linear\,\cite{cklaw}, approximate models are usually used \cite{Aspelmeyer,optobook} which assume a linear dependence of the cavity frequency $\omega(X_b)$ on the dimensionless position of the movable mirror, $X_b$. So far these ``linear" models have proved extremely successful, aided by the fact that the bare (or `single-photon') optomechanical coupling strength is usually very small\,\cite{Aspelmeyer,optobook}. However, researchers are continuously exploring ways of enhancing the optomechanical coupling, as well as the potential of optomechanics for ultra-high-accuracy applications such as Planck physics \cite{s-kumar}. Hence, extensions to the linear model are becoming a necessity. The next step beyond the linear approach is to expand the cavity frequency up to and including second order in $X_b$, leading to what we will call the ``quadratic model" \,\cite{sala}.
\begin{figure}[t]
\begin{center}
  \includegraphics[width=\linewidth]{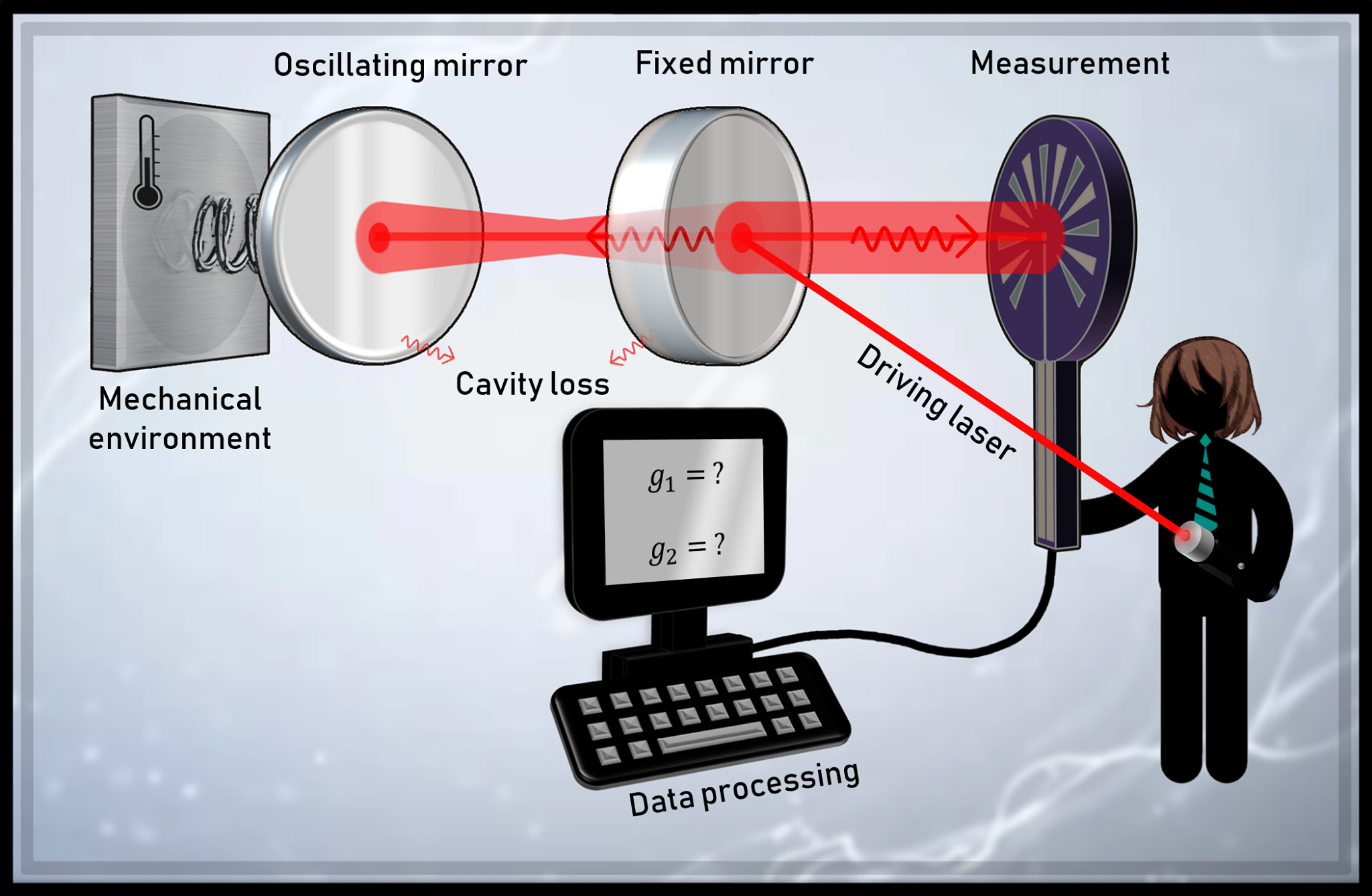}  
\end{center}
\caption{Schematic of the parameter estimation methodology for driven-dissipative optomechanics. We consider a driven-dissipative optomechanical system featuring a driven (by an external laser) and lossy (photons escaping the cavity) cavity and a damped mechanical oscillator. The mechanical support has low but finite temperature (leading to a non-zero thermal occupation number). The optomechanical coupling arises due to the radiation pressure on the movable mirror. Once the system has reached a steady state we measure an observable. We repeat the measurement many times to get the statistics. Finally, we process the data to find the best guess for the coupling parameters of interest.}
\label{fig:0}
\end{figure}
Accurate knowledge of all the relevant optomechanical coupling parameters will indeed be crucial for virtually any application of these systems. With such motivation in mind, this paper exploits local quantum estimation theory\,\cite{paris} (QET) to investigate how precisely the linear and quadratic coupling strengths may be measured in a model quantum optomechanical system. In a nutshell, QET looks for the best strategy for estimating unknown parameters encoded in the density matrix of a quantum system (i.e. a \textit{quantum statistical model}) \,\cite{estimation,estimationnopto}. 
The ultimate limits to the precision with which the desired parameters can be estimated may be quantified via the quantum Cram\'{e}r-Rao bounds \cite{paris,safranek,multiparameter}.

Within such framework, we consider an optomechanical set-up featuring a driven (and lossy) cavity, whose dynamics is described by a Lindblad master equation. As is typical in recent optomechanics experiments, we assume sufficiently strong driving to approximate the dynamics via a master equation that is bilinear in the canonical operators. This leads to a Gaussian steady state \cite{gaussianstates}, whose first and second moments we characterise via a combination of matrix algebra and numerical methods. This state, with its explicit dependence on all the model parameters (in particular the unknown coupling strengths), will embody our quantum statistical model. This in turn may be attacked via general closed-form expressions that are available for QET in Gaussian models\,\cite{gerardo}. The visual representation of the parameter estimation methodology for driven-dissipative optomechanics is shown in Fig. \ref{fig:0}.

Using this approach, and for typical values of the model parameters (inspired by recent experiments), we found that it is significantly easier to estimate the linear coupling strength than the quadratic one. This result is indeed to be expected in typical experimental conditions, given the significant difference in the relative strength of the two constants. Our analysis also reveals that the majority of information about the parameters is encoded in the reduced state of the mechanical element. 

We then investigate how well some specific measurements perform when compared to the fundamental limits imposed by QET. We in particular focus on measurements of the mechanical position $X_b$, field amplitude $Q$, mechanical momentum $P_b$ and field phase $P$, which can all be analyzed within the Gaussian formalism \cite{serafini}. Among those we find that the best strategy to estimating the coupling parameters is through a direct measurement of the mechanical position $X_b$. 

We additionally explored the influence of temperature on the estimation precision of the coupling strengths. In the case of the linear coupling parameter we found that the effect of temperature is mostly significant at lower intracavity photon numbers (i.e. lower driving strengths), where it improves the estimation precision. At higher intracavity photon numbers, the zero temperature scenario predicts a better estimation precision instead. Interestingly, in the case of the quadratic coupling parameter we found that a hotter mechanical bath gives a higher estimation precision at all intracavity photon numbers in our range. 

We note that the application of QET in closely related optomechanical set-ups was previously 
considered in works by Bernad, Sanavio and Xuereb\,\cite{estimationnopto,sanavio}, who adopted the linear model of
optomechanics. In Ref.~\cite{estimationnopto}, purely Hamiltonian (non-dissipative) dynamics was assumed, and it was found that larger intracavity photon numbers would facilitate the estimation of the linear coupling strength. Analogously, our results to follow show that a similar conclusion holds in the strong driving regime. However, we also find that for weaker drive strengths the picture is more complicated when considering finite-temperature effects. We note that very recently the same authors went on to consider a driven-damped system~\cite{sanavio} though using a somewhat different approach to ours and without considering quadratic couplings. In particular, Ref.~\cite{sanavio} neglects the contribution of the steady state's first moments (i.e. the averages of the canonical operators) to the quantum Fisher information (QFI). We checked that this is a well-justified assumption for the model parameters adopted therein. However, our results show clearly that there are experimentally accessible parameter regimes where the picture changes dramatically and the first moments can come to dominate the QFI. Finally, Schneiter et al.~\cite{nonlinearregime} have applied local QET to a time-dependent, purely Hamiltonian and quadratic optomechanical system. Such a model, however, is sufficiently different from ours to prevent a simple and direct comparison between the two.

This paper is organised as follows. In Sec.~\ref{model} we introduce our model of driven-dissipative optomechanics, including both linear and quadratic coupling terms, and outline how the dynamics may be approximated via a bilinear master equation in the limit of strong driving. In Sec.~\ref{theoreticalmethods} we calculate the steady state of the system, which is Gaussian within the considered approximations, and hence is fully characterized by its first and second moments. We then develop the necessary QET tools to investigate the optimal estimation of linear and quadratic coupling parameters. In Sec.~\ref{results} we present and discuss the findings of our research. Finally, in Sec.~\ref{summary} we summarise our results.
\section{Model}
\label{model}
\label{sec:optomodels}
We consider a simple optomechanical system consisting of two quantum harmonic oscillators, describing a single-mode cavity field and a single mechanical mode, respectively. The two modes are coupled non-linearly via radiation pressure \cite{Aspelmeyer,optobook,backaction}. We thus assume that our system is described by the Hamiltonian
\begin{align}
H_0&=\hbar\omega(\hat X_b)\frac{\hat Q^2+\hat P^2}{2}+\hbar\omega_m\frac{\hat X_b^2+\hat P_b^2}{2},
\end{align}
with $\hat Q$ and $\hat P$ the amplitude and phase quadratures for the cavity mode, $\hat X_b$ and $\hat P_b$ the dimensionless position and momentum operators of the movable mirror, whilst $m$ and $\omega_m$ are the effective mass and frequency of the mechanics \cite{sala,estimationnopto,cklaw,operator}. The only nontrivial commutators read $[\hat Q,\hat P]=[\hat X_b,\hat P_b]=i$. The optomechanical coupling arises from a parametric dependence of the cavity frequency on the mechanical position $\omega(\hat X_b)$.

In the widely explored ``linear" regime of optomechanics, the mechanical motion is assumed to be very small and $\omega(\hat X_b)$ is approximated by an expansion to linear order in $\hat X_b$. For the ``quadratic model" of optomechanics, instead, terms up to and including $\hat X_b^2$ are retained: 
\begin{align}
\omega(\hat X_b)\approx \omega_0+\omega'(0) \hat X_b+\frac{1}{2} \omega''(0)\hat X_b^2,
\end{align}
where $\omega_0$ is the bare cavity frequency \cite{Aspelmeyer,optobook,sala}. The strength of the optomechanical interaction can be quantified with the linear and quadratic coupling strengths, which for a generic set-up are defined as
\begin{align}
    g_1&\equiv \frac{1}{\sqrt{2}}\omega'(0),\label{g1def}\\ g_2&\equiv \frac{1}{2}\omega''(0),\label{g2def}
\end{align}
 respectively \cite{optobook}. Note that we can always ensure that $g_1$ is positive by a redefinition of the positive direction of $X_b$, and that the linear model is recovered by setting $g_2$ to zero.

A purely Hamiltonian description of the system is however not sufficient for our purposes, since we aim to describe a (more realistic) driven-dissipative optomechanical system featuring a driven and lossy cavity, and a damped mechanical oscillator. In order to conveniently introduce coherent driving in the model, we shall move to a frame rotating at the frequency of the driving laser, $\omega_L$. In this frame the Hamiltonian of the driven system may be written as
\begin{align}
H  = &\hbar\left(\Delta_0+\omega'(0) \hat X_b+\frac{1}{2} \omega''(0)\hat X_b^2\right)\frac{\hat Q^2+ \hat P^2}{2}\nonumber\\
&+\hbar\omega_m\frac{\hat X_b^2+\hat P_b^2}{2}+\sqrt{2}\hbar{\cal E} \hat Q,
\end{align}
where $\Delta_0=\omega_0-\omega_L$ is the detuning between the cavity and driving laser and ${\cal E}$ is the drive amplitude. 

In our model we will include cavity decay at a rate $\kappa$ and mechanical damping at a rate $\Gamma_m$, assuming that the thermal occupation of the cavity mode is negligible. We assume that the corresponding master equation describing the dynamics of the system is of the general Lindblad form \cite{easy,breuer}:
\begin{align}
\label{genformme}
    \dot{\rho}(t)=&-\frac{i}{\hbar}[H,\rho(t)]\nonumber\\
    &+\sum_{ij}\frac{\gamma_{ij}}{2}\left[2\hat R_i\rho(t)\hat R_j-\{\hat R_j\hat R_i,\rho(t)\}\right],
\end{align}
where we defined the vector of quadrature operators 
\begin{align}
\hat{\boldsymbol{R}}&=(\hat Q,\hat P,\hat X_b,\hat P_b),\label{Rdef}
\end{align}
while $\boldsymbol{\gamma}$ is the damping matrix:
\begin{align}
\label{dampingmatrix}
    \boldsymbol{\gamma}=
    \begin{pmatrix}
    \frac{\kappa}{2} &-i\frac{\kappa}{2} &0 & 0\\
i\frac{\kappa}{2} & \frac{\kappa}{2} &0 &0\\
0& 0& \frac{\Gamma_m}{2}(2\bar{n}_m+1)& -i\frac{\Gamma_m}{2}\\
0& 0& i\frac{\Gamma_m}{2}&\frac{\Gamma_m}{2}(2\bar{n}_m+1)
    \end{pmatrix},
\end{align}
$\bar{n}_m=1/\left(e^{\hbar\omega_m/k_BT}-1\right)$ is the mean occupancy of the mechanical oscillator, $k_B$ is the Boltzmann constant and $T$ is the temperature of the mechanical reservoir \cite{Aspelmeyer,optobook}. We note that, in choosing a Lindblad form, we automatically excluded the use of the standard Brownian motion master equation (SBMME) \cite{breuer} to describe mechanical damping. Indeed, a Lindblad form greatly simplifies our analysis, since it avoids non-positivity issues that are known to occur in the SBMME \cite{vitali}.

The main effect of the drive is to displace the steady states of both the cavity field and the mechanical position \cite{optobook}. We assume that the cavity is driven sufficiently strongly (and that the optomechanical couplings are weak enough) so that the system dynamics can be approximated via a bilinear master equation description, where only small quantum fluctuations around the semi-classical steady state are considered \cite{optobook,sanavio}. In detail, we start by displacing our canonical operators as per $\hat{\mathbf{R}}\to\hat{\mathbf{R}}+\mathbf{R}_0$, where 
\begin{align}
  \mathbf{R}_0&=(Q_0,P_0,x_0,p_0) \label{averages} 
\end{align} 
is the vector of steady-state quadrature averages. Here, $x_0$ and $p_0$ are the average position and momentum of the mechanics in the steady state, while $Q_0$ and $P_0$ are the steady state displacements of the amplitude and phase quadratures, respectively \cite{optobook}. Of course, the steady-state expectation values of the transformed operators will now vanish. 
This results in the following equations for the steady-state values of the system's first moments:
\begin{align}
\label{average1}
Q_0 &=\frac{-2\Delta_{eff}\mathcal{E}}{\sqrt{2}\left(\Delta_{eff}^2+\frac{\kappa^2}{4}\right)},\\
\label{average2}
P_0&=\frac{-\kappa\mathcal{E}}{\sqrt{2}\left(\Delta_{eff}^2+\frac{\kappa^2}{4}\right)},\\
\label{average3}
x_0&=-\frac{\omega'(0)\omega_m\mathcal{E}^2}{\left(\omega_m^2+\frac{\Gamma_m^2}{4}\right)\left(\Delta_{eff}^2+\frac{\kappa^2}{4}\right)+\omega''(0)\omega_m\mathcal{E}^2},\\
\label{average4}
p_0&=\frac{\Gamma_m}{2\omega_m}x_0,
\end{align}
where $\Delta_{eff}=\Delta_0-\sqrt{2}g_1x_0+g_2x_0^2$ is the effective detuning. The non-linearity of Eqs. (\ref{average1}-\ref{average4}) suggests that multiple steady state solutions are possible \cite{Aspelmeyer,optobook}. This is known as dynamical multistability \cite{Aspelmeyer,optobook}. In detail, depending on the driving strength up to five (quadratic model) or three (linear model) different steady states solutions can exist. In this work, we only focus on parameters regimes where the system is \textit{stable}, i.e., where a unique real solution to Eqs.~(\ref{average1}-\ref{average4}) exists. This in turn places an upper bound to the drive strength $|\mathcal{E}|^2$ \cite{estimationnopto}. 

After the displacement has been implemented, we neglect terms that are beyond quadratic order in the transformed canonical operators \cite{optobook}. The corresponding master equation reads
\begin{align}
\label{linearisedquadraticrho}
    \dot{\rho}(t)=&-\frac{i}{\hbar}[H_B,\rho(t)]\nonumber\\
    &+\sum_{ij}\frac{\gamma_{ij}}{2}\left[2\hat R_i\rho(t)\hat R_j-\{\hat R_j\hat R_i,\rho(t)\}\right],
\end{align}
where we note that the Lindblad operators remain unchanged, while the Hamiltonian now takes the bilinear form
\begin{align}
\label{linearisedham}
     H_B=&\frac{\hbar}{2}\Delta_{eff}\left(\hat Q^2+\hat P^2\right)+\frac{\hbar}{2}\omega_m \hat P_b^2+\frac{\hbar}{2}\omega_{eff}\hat X_b^2\nonumber\\
     &+\hbar g_{eff}\hat X_b\left(\hat QQ_0+\hat PP_0\right),
\end{align}
with $\omega_{eff}=\omega_m+2g_2|\alpha|^2$ the effective mechanical frequency and $g_{eff}=-\sqrt{2}g_1+2g_2x_0$ the effective coupling strength. Note that the assumption of strong cavity driving translates into the condition that the intracavity photon number is large, i.e., $|\alpha|^2\equiv(Q_0^2+P_0^2)/2\gg 1$.

\section{Estimating coupling constants from the steady state}
\label{theoreticalmethods}
\subsection{Covariance Matrix Formalism}
Due to its bilinear form, the master equation \eqref{linearisedquadraticrho} admits a Gaussian steady state that can, in general, be fully characterised by its first and second moments of the quadrature operators $\hat{\mathbf{R}}$ \cite{gaussianstates}. After having determined our steady state, we will be able to exploit general closed-form expressions that are available for QET in Gaussian models \cite{gerardo}.

As anticipated in the previous section, the first moments of our Gaussian steady state are given by $\mathbf{R}_0=(Q_0,P_0,x_0,p_0),$ and are found by solving Eqs.~(\ref{average1}-\ref{average4}) --- recall also that we will only consider parameter regimes in which such solution is unique. The second moments are instead encoded in the steady state covariance matrix $\bar{\sigma}$, which in our displaced frame of reference is given by
\cite{gaussianstates}
\begin{align}
\label{covariancematrix}
\scalemath{0.8}{
{\boldsymbol{\bar{\sigma}}}=\begin{bmatrix}
\langle \hat Q^2 \rangle&\langle \frac{1}{2}\{\hat Q,\hat P\}\rangle&\langle \hat Q\hat X_b\rangle &\langle \hat Q\hat P_b\rangle\\
\langle \frac{1}{2}\{\hat Q,\hat P\}\rangle&\langle \hat P^2\rangle&\langle \hat P\hat X_b\rangle&\langle \hat P\hat P_b\rangle\\
\langle \hat X_b\hat Q\rangle&\langle \hat X_b\hat P\rangle &\langle \hat X_b^2\rangle &\langle \frac{1}{2}\{\hat X_b,\hat P_b\}\rangle\\
\langle \hat P_b\hat Q\rangle&\langle \hat P_b\hat P\rangle&\langle \frac{1}{2}\{\hat X_b,\hat P_b\}\rangle&\langle \hat P_b^2\rangle
\end{bmatrix}
},
\end{align}
where $\{\hat A,\hat B\}\equiv \hat A\hat B+\hat B\hat A$ is the anticommutator. As detailed in Appendix~\ref{covariancematrixlanguage}, master equation \eqref{linearisedquadraticrho} implies the following Lyapunov equation for the steady state covariance matrix:
\begin{align}
\label{Lyapunovsolve1}
    B^T{\boldsymbol{\bar{\sigma}}}+{\boldsymbol{\bar{\sigma}}}B=C,
\end{align}
where
\begin{align}
\label{B}
    B&=\frac{i}{\hbar}\boldsymbol{HW}+\boldsymbol{\gamma_AW},\\
    C&=-\boldsymbol{W\gamma_SW},\\
    \boldsymbol{H}&=
    \begin{pmatrix}
    \hbar\Delta_{eff}& 0& \hbar g_{eff} Q_0& 0\\
    0&\hbar\Delta_{eff} & \hbar g_{eff}P_0& 0\\
    \hbar g_{eff}Q_0&\hbar g_{eff}P_0 &\hbar\omega_{eff}& 0\\
    0& 0& 0& \hbar\omega_m
    \end{pmatrix},\\
    \boldsymbol{W}&=
    \begin{pmatrix}
0 & i &0 & 0\\
-i & 0 &0 &0\\
0& 0& 0& i\\
0& 0& -i&0
\end{pmatrix},\\
    \boldsymbol{\gamma_S}&=\frac{\boldsymbol{\gamma}+\boldsymbol{\gamma}^T}{2},\\
    \boldsymbol{\gamma_A}&=\frac{\boldsymbol{\gamma}-\boldsymbol{\gamma}^T}{2}.
\end{align}
We note that Eq.~\eqref{Lyapunovsolve1} can be solved analytically in terms of the model parameters and the vector of averages $\mathbf{R}_0$. The latter, however, may in general not admit an analytical expression in terms of the model parameters, as we recall it is the solution to the nonlinear system of equations \eqref{average1}-\eqref{average4}.
In the next section we shall show how to develop a comprehensive QET analysis of the coupling parameters solely from the knowledge of the first and second moments of our Gaussian steady state.

\subsection{Quantum Estimation Theory for Gaussian States}
The aim of quantum estimation theory (QET) is to identify the best strategy for estimating one or more parameters encoded in the density matrix of a quantum system \cite{paris,estimation,estimationnopto}. Here we focus on local QET, which seeks a strategy that maximises the Fisher information over all possible measurements, and implicitly assumes that a rough estimate of the parameter value is known in advance \cite{paris}. 

In our model of driven-dissipative optomechanics, the parameters to be estimated shall be the coupling strengths $g_1$ and $g_2$. As anticipated, all of the information about these parameters will be contained in the steady-state averages, $\mathbf{R}_0$, as well as in the steady state covariance matrix, $\boldsymbol{\bar{\sigma}}$. Specifically, for our coupling parameters $(g_1,g_2)$ the elements of the quantum Fisher information matrix (QFIM) are given by
\begin{align}
\label{QFIM}
    I_{i,j}&=\left(\partial_{g_i}\boldsymbol{R}_0^T\right)\boldsymbol{\bar{\sigma}}^{-1}\left(\partial_{g_j}\boldsymbol{R}_0\right)\nonumber\\
    &+2Tr\left[\left(\partial_{g_i}\boldsymbol{\bar{\sigma}}\right)\left(4\mathcal{L}_{\boldsymbol{\bar{\sigma}}}+\mathcal{L}_W\right)^{-1}(\partial_{g_j}\boldsymbol{\bar{\sigma}})\right], 
\end{align}
where $\mathcal{L}_{\boldsymbol{\bar{\sigma}}} (\boldsymbol{A})=\boldsymbol{\bar{\sigma}} \boldsymbol{A}\boldsymbol{\bar{\sigma}}$, $\mathcal{L}_W (\boldsymbol{A})=\boldsymbol{W} \boldsymbol{A}\boldsymbol{W}$. Note also that the term $\left(4\mathcal{L}_{\boldsymbol{\bar{\sigma}}}+\mathcal{L}_W\right)^{-1}$ refers to the pseudoinverse if the term inside the bracket is singular \cite{gerardo,monras}. The first term is the contribution due to the averages, while the second term is the contribution due to the variances and covariances towards the total QFI \cite{monras}. This terminology will be convenient later on as we seek to unravel how the different terms contribute across different parameter regimes. We note, however, that our terminology only describes the origin of the dependence of the gradients with respect to the coupling parameters. Hence whilst the first term in  eqn.\ (\ref{QFIM}) only contains gradients of the averages  with respect to the coupling constants, and we will therefore call it the contribution of the averages, it does also depend on the covariance matrix. Eq. (\ref{QFIM}) facilitates efficient numerical computation of the QFI \cite{monras}. 

The ultimate limit to parameter estimation in this context is set by the QCRB \cite{paris,safranek,multiparameter}. In multi-parameter estimation theory both coupling parameters are assumed to be unknown (or only known with low precision). In this case, the QCRB relates the covariance matrix of any pair of unbiased estimators for the parameters $(g_1, g_2)$ to the QFIM. For $M$ experimental runs, the corresponding QCRB reads \cite{multiparameter,gerardo}
\begin{align}
Cov(g_1,g_2)\geq \frac{1}{M} \boldsymbol{I}^{-1}.
\end{align}
The limiting case of single-parameter estimation theory can be reached if we assume that only one parameter is unknown, say $g_i$. In this case the QCRB relates the variance $Var(g_i)$ of an unbiased estimator of the parameter $g_i$ to the corresponding diagonal element of the QFIM. For $M$ experimental runs, the corresponding bound reads \cite{gerardo,serafini,monras,zwierz,zou,abinitio}
\begin{align}
\label{QCRB}
    Var(g_i) \geq \frac{1}{M\,I_{ii}}.
\end{align}
In other words, the diagonal elements of the QFI matrix quantify the ``best-case-scenario" performance for the estimation of each individual parameter. Hence, in what follows we shall pick Eq.~\eqref{QCRB} as our benchmark in evaluating the performance of various measurements (see below). Note that in single-parameter estimation theory the saturation of the QCRB is guaranteed, at least in the limit $M\to \infty$, and assuming that every mathematically allowed quantum measurement can be implemented \cite{multiparameter,gerardo,monras}. This, however, is not true for the estimation of multiple parameters: in this case the optimal measurements for different parameters may not be compatible \cite{multiparameter}. 

While the QFI quantifies the ultimate quantum limit to parameter estimation \cite{paris,multiparameter,advanced}, the estimation performance of specific measurement strategies may be quantified via the
classical Fisher information (FI) matrix \cite{paris,luati}. In our context, the FI measures the amount of information that a classical random variable $s$ (the outcome of a quantum measurement) contains about the parameters $g_1,g_2$ \cite{fisher}. The FI matrix elements take the form
\begin{align}
\label{classicalfisher}
J_{i,j}=\int_{-\infty}^{\infty} ds \frac{1}{p_{g_1,g_2}(s)}\left(\frac{\partial p_{g_1,g_2}(s)}{\partial g_i}\right)\left(\frac{\partial p_{g_1,g_2}(s)}{\partial g_j}\right),
\end{align}
where $p_{g_1,g_2}(s)$ is the probability distribution of the measurement outcome $s$, assumed to be a smooth function of $(g_1,g_2)$ \cite{gerardo,fisher}. Depending on the chosen observables, analytical solutions to the integral in Eq. (\ref{classicalfisher}) may exist. This is particularly true for quadrature measurements (i.e. a measurement of $\hat Q, \hat P, \hat X_b, \hat P_b$ or a linear combination thereof), provided that the measured state is Gaussian \cite{serafini}.

In the case of optomechanics, it is well known that one can use a homodyne detection scheme to measure the light quadratures $\hat Q$, $\hat P$ \cite{homodyne,serafini}. However, we shall also consider a direct measurement of the mechanical quadratures, $\hat X_b$ and $\hat P_b$, for completeness. In practice this could potentially be achieved using e.g. another optical mode of the cavity. In this scenario, the probability distribution associated with a measurement of $\hat s\in\{\hat Q,\hat P,\hat X_b,\hat P_b\}$   has the following expression \cite{serafini}:
\begin{align}
p_{g_1,g_2}(s)=\frac{e^{-\frac{(s-s_0(g_1,g_2))^2}{2\bar\sigma_{kk}(g_1,g_2)}}}{\sqrt{2\pi\bar{\sigma}_{kk}(g_1,g_2)}},
\end{align}
where $s_0(g_1,g_2)$ is the steady state average of the chosen quadrature, appropriately chosen from the set $\{Q_0,P_0,x_0,p_0\}$, while $\bar{\sigma}_{kk}(g_1,g_2)$ is the corresponding diagonal element of the steady state covariance matrix ($\bar{\sigma}_{11}$ for $\hat s=\hat Q$, $\bar{\sigma}_{22}$ for $\hat s=\hat P$ and so on). In this setting an analytical solution to the integral in Eq. (\ref{classicalfisher}) exists and is given by 
\begin{align}
\label{analyticalFI}
J_{i,j}&=\frac{1}{2\bar{\sigma}_{kk}(g_1,g_2)^2}\times\nonumber\\
&\Big[2\bar{\sigma}_{kk}(g_1,g_2)\left(\frac{\partial s_0(g_1,g_2)}{\partial g_j}\right)\left(\frac{\partial s_0(g_1,g_2)}{\partial g_i}\right)\nonumber\\
&+\left(\frac{\partial \bar{\sigma}_{kk}(g_1,g_2)}{\partial g_j}\right)\left(\frac{\partial \bar{\sigma}_{kk}(g_1,g_2)}{\partial g_i}\right)\Big].
\end{align}
Note that the choice of a strategy to estimate the parameters $(g_1,g_2)$ is optimal if the FI and QFI matrices are equal, i.e. $\boldsymbol{J}=\boldsymbol{I}$.

As anticipated, we shall focus solely on the diagonal elements of the QFIM (Eq. (\ref{QFIM})): $I_{1,1}$ and $I_{2,2}$, respectively. As noted above, the diagonal elements are indeed the ``most optimistic'' quantifiers of estimation precision of the coupling strengths. In general, however, the combined precision of the two parameter estimations will be worse than what the diagonal elements suggest.  
 
Both, the definitions of the QFI (Eq. (\ref{QFIM})) and the FI (Eq. (\ref{analyticalFI})) rely on the derivatives of the steady state covariance matrix and the averages with respect to the coupling strengths. Since we have seen that both $\bar{\sigma}$ and $\mathbf{R}_0$ are determined by the nonlinear system of equations \eqref{average1}-\eqref{average4}, which in general can only be solved numerically, we use implicit differentiation to calculate the derivatives in question. This allows us to express all our quantities of interest in terms of the numerical solution to the above nonlinear equations, and allows us to avoid numerical differentiation altogether. 
\begin{figure*}[t!]
\begin{center}
         \includegraphics[width=.475\linewidth]{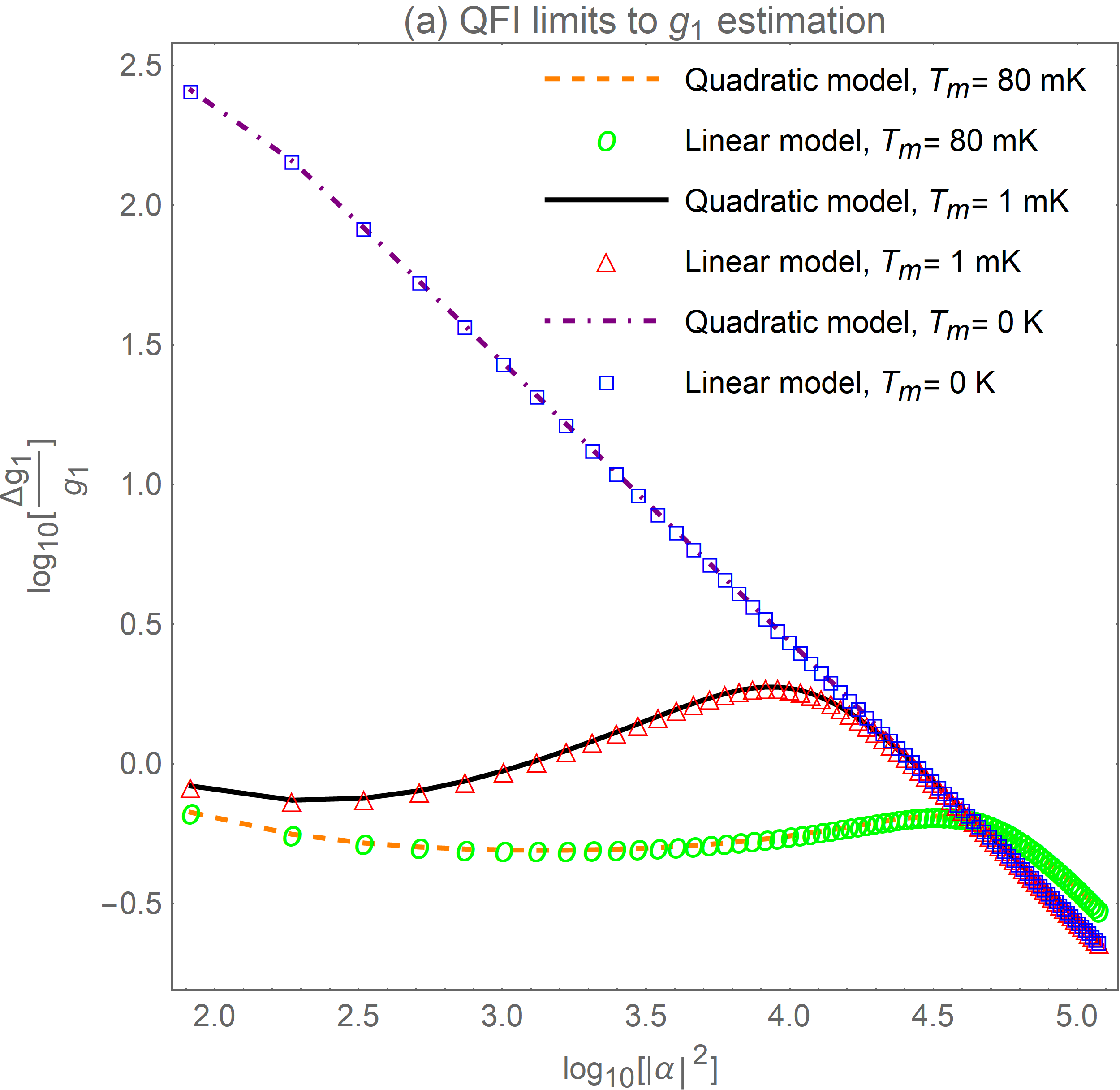}
         \hfill
         \includegraphics[width=.475\linewidth]{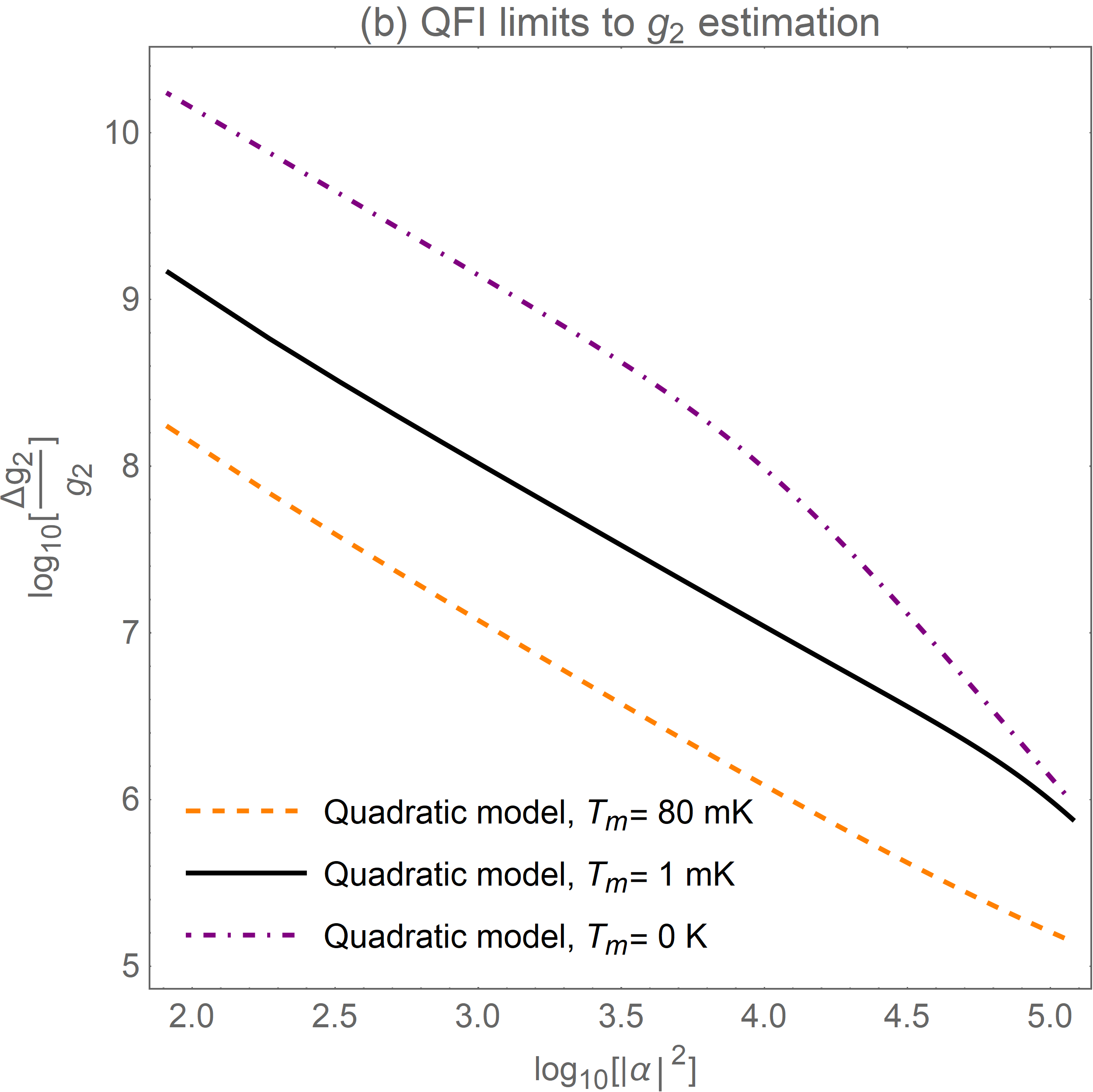}
         \end{center}
     \caption{{(a)} Log-log plot of the relative error bound on $g_1$ (as implied by QFI) against the intracavity photon number, $|\alpha|^2$ as predicted by the linear and quadratic models in the zero temperature (green circles and orange dashed line, respectively), low temperature (red triangles and black line, respectively) and high temperature (blue squares and purple dot-dashed line, respectively) scenarios. {(b)} Log-log plot of the relative error bound on $g_2$ (as implied by QFI) against the intracavity photon number, $|\alpha|^2$ as predicted by the quadratic model in the zero temperature (purple dot-dashed line), low temperature (black line) and high temperature (orange dashed line) scenarios.}
     \label{Fig1}
\end{figure*}
\section{Results}
\label{results}
For simplicity we consider a specific geometry, that of a Fabry-Perot cavity \cite{Aspelmeyer}, in which one mirror is fixed and the other is mounted on the mechanical oscillator. In this case, assuming an ideal one-dimensional cavity field, the cavity frequency takes the specific form
\begin{align}
    \omega(\hat X_b)=\frac{\omega_0}{1+\frac{\sqrt{2}x_{zp}\hat X_b}{L}},
\end{align}
with $L$ the bare cavity length and $x_{zp}=\sqrt{\hbar/2m\omega_m}$ the ground state position uncertainty of the mechanical oscillator \cite{optobook}. Making the standard assumption that the mechanical motion is very small on the scale of the cavity length $\hat X_b/L\ll 1$, for the quadratic model of optomechanics $\omega(\hat X_b)$ is approximated as 
\begin{align}
    \omega(\hat X_b)\approx \omega_0-\sqrt{2}g_1\hat X_b +g_2 \hat X_b^2,
\end{align}
where $g_1=\omega_0 x_{zp}/L$ and $g_2=2\omega_0 x_{zp}^2/L^2$ are the linear and quadratic coupling strengths in accordance with Eqs. (\ref{g1def}) and (\ref{g2def}), respectively \cite{Aspelmeyer,optobook,sala}. 

\begin{figure*}[t]
     \centering
         \includegraphics[width=.475\linewidth]{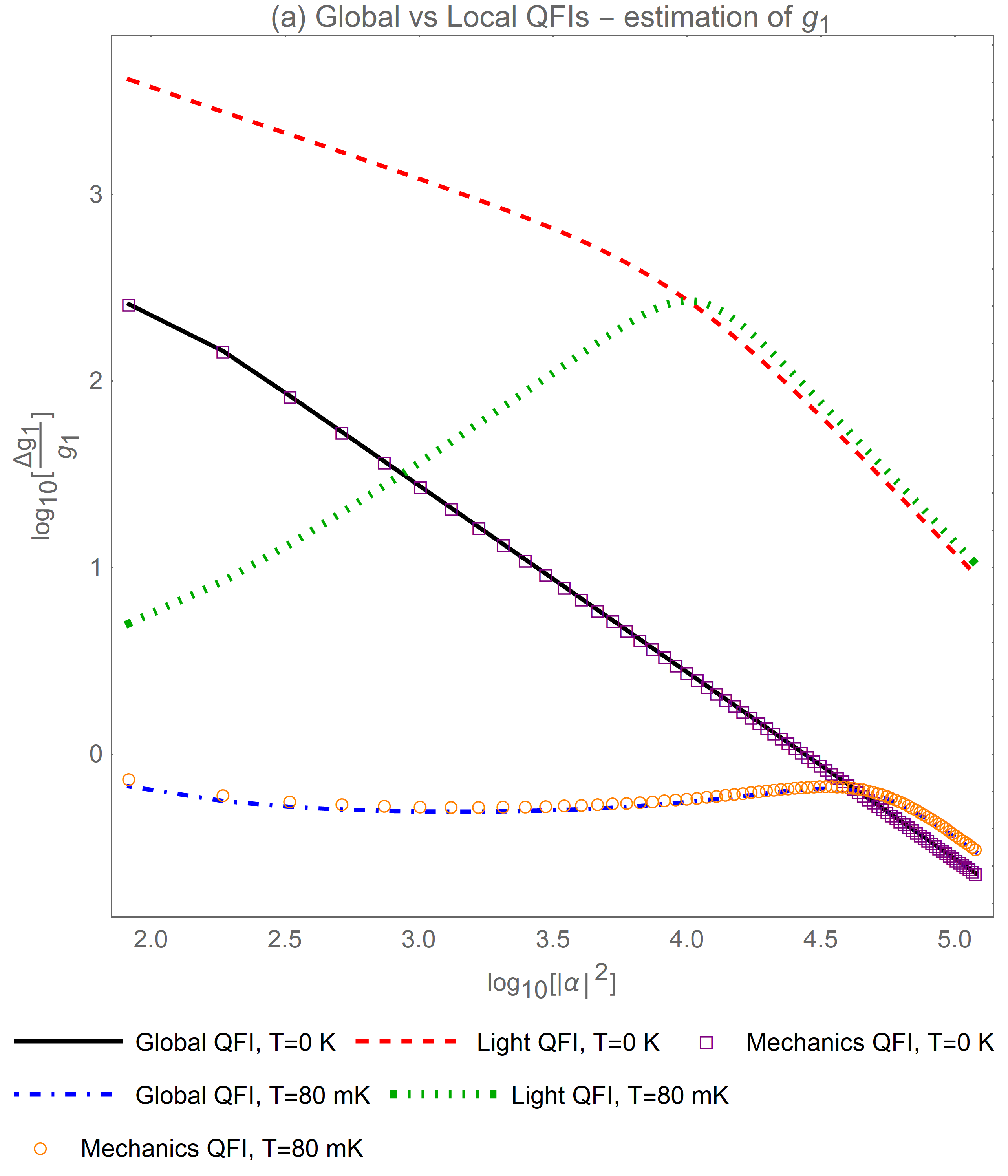}
     \hfill
         \includegraphics[width=.475\linewidth]{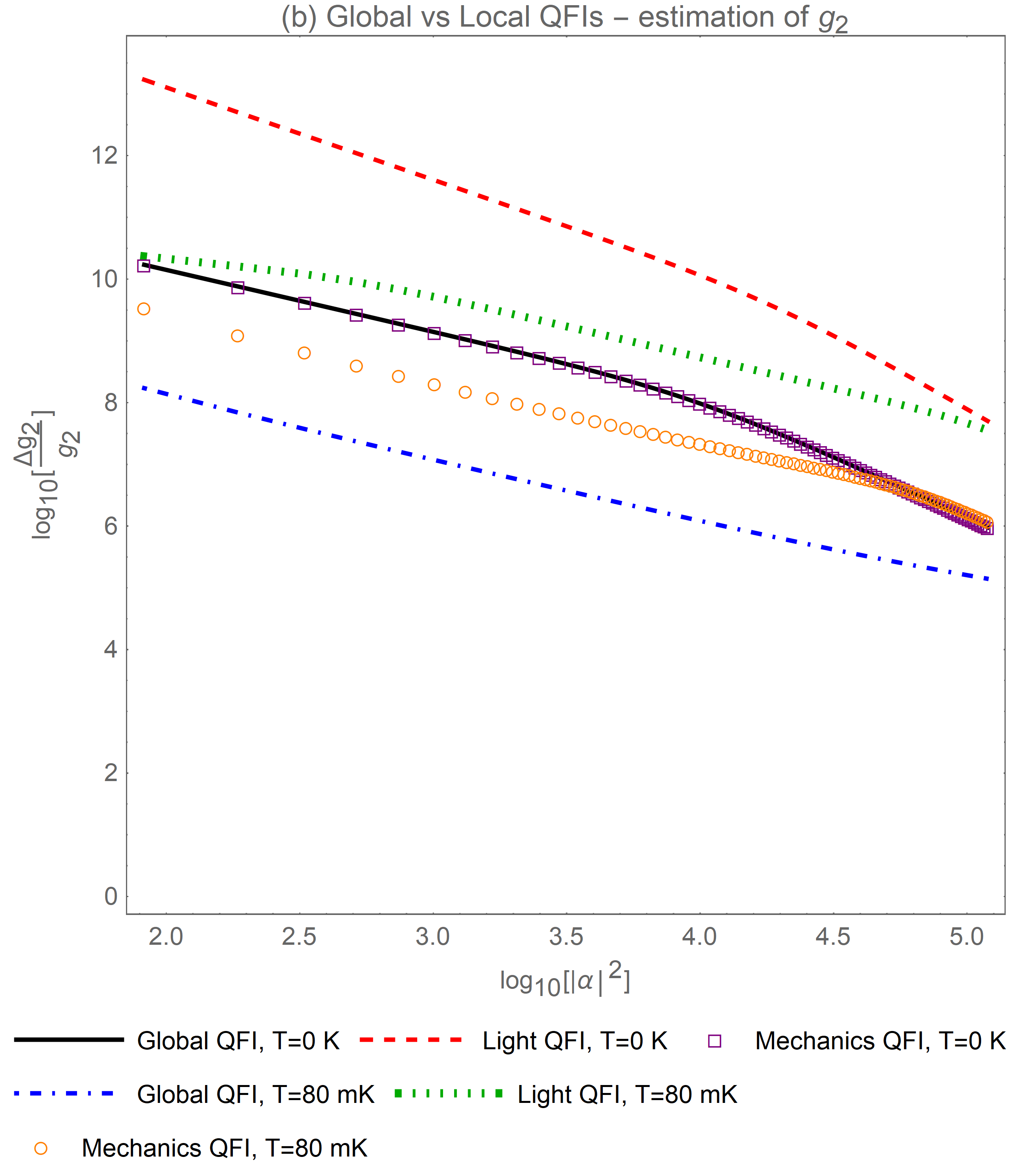}
     \caption{Relative error bounds on coupling constants (a) $g_1$ and (b) $g_2$, as a function of the intracavity photon number, $|\alpha|^2$ as implied by the global, light and mechanics QFIs in the case of zero temperature (black line, red dashed line, purple squares, respectively) and high temperature (blue dot-dashed line, green dotted line and orange circles, respectively).}
     \label{Fig2}
\end{figure*}

Here, we examine three scenarios: zero temperature ($T=0$ K), low temperature ($T=1$ mK) and ``high'' temperature ($T=80$ mK) scenario. In each case we are looking for the best strategy to estimate the linear, ${g}_1$, and quadratic, ${g}_2$, coupling strengths. First, we establish the fundamental quantum limits on the estimation precision, which, in accordance with the quantum Cram\'{e}r-Rao bound (QCRB), are quantified with the ``global" QFIs (i.e. the QFIs calculated from the bipartite state of light plus mechanics). Additionally, by tracing out the mechanical (light) mode, we can calculate ``local" QFIs that are relevant when only the light (mechanical) mode is directly measurable. Comparing these local QFIs with the global ones will also reveal how much information about the coupling parameters is contained in the reduced states of light and mechanics. Finally, we compare the QFI limits to the performance of a small selection of ``realistic'' measurements (quantified with the respective FI), including those of $\hat Q$, $\hat P$, $\hat X_b$ and $\hat P_b$. This can help us discern which of the experimentally common measurements constitute the best strategy to parameter estimation in each scenario. In many cases we have chosen to measure the estimation precision with the relative error 
\begin{equation}
\frac{\Delta g_i}{g_i}\geq\frac1{g_i\sqrt{I_{i,i}}}. 
\end{equation}

Our choice of parameter values is motivated by recent experiments where the ground state of a mechanical oscillator was approached via back-action cooling arising from a red-detuned laser drive, specifically \cite{goodcavity}. Correspondingly, we adopt the following parameter values $\omega_m=1.1\times 10^7$ Hz, $m=4.8\times 10^{-14}$ kg, $\Gamma_m=32$ Hz, $\Delta_0=\omega_m$, $\kappa=10^5$ Hz, $g_1=2\times 10^2$ Hz and $g_2=2g_1^2/\omega_0\approx 1.1\times 10^{-5}$ Hz \cite{goodcavity}. In order to ensure that the driving is strong enough for the Gaussian approximation to hold and we do not encounter any stability issues we consider a region $10^8\leq\mathcal{E}\leq3.8\times10^9$ Hz in all three scenarios. In terms of the intracavity photon number, $|\alpha|^2$, this corresponds to a region: $80\lesssim|\alpha|^2\lesssim1.2\times 10^5$ (or $1.9 \lesssim \log_{10}(\mid\alpha\mid^2)\lesssim 5.1$).

In Fig. \ref{Fig1}(a) we investigate the effects of the higher order $g_2$ term, temperature and driving on the estimation precision of the linear coupling strength, $g_1$. Clearly, in all scenarios the effect of the higher order corrections due to the $g_2$ term is minimal: the linear and quadratic models show a very good agreement at all $|\alpha|^2$. This is to be expected as $g_2$ is several orders of magnitude below $g_1$ in our example. As an interesting aside, in a membrane-in-the-middle optomechanical system it is possible to engineer a purely quadratic coupling via the position of the membrane, in which case $g_2$ would clearly play the key role. \cite{optobook,membrane}. 

Figure \ref{Fig1}(a) reveals a surprisingly complex dependence on temperature. There is a crossover around  $\log_{10}(|\alpha|^2)\sim 4.6$ (or $|\alpha|^2\sim 4\times 10^4$): below this value the high temperature scenario offers the best precision for estimating $g_1$, but above it the best precision is found at lower temperatures. As discussed in Appendix \ref{Appendix} (see in particular Fig. \ref{fig:4}), much of this behavior can be understood by looking at the relative contributions of the variances and averages to the QFI and how these change with temperature. The contribution of the averages to the QFI always increases monotonically with the intracavity photon number and hence it always eventually dominates. This, taken together with the fact that the contribution of the averages to the QFI is reduced by increasing the temperature of the mechanical reservoir, means that the best precision is eventually expected at zero-temperature. However, for non-zero  temperatures the contribution to the QFI from the variances is important and it is dominant at sufficiently low intracavity photon numbers. This is not surprising given the strong cooling effect that the driven cavity can have on the mechanics\,\cite{Aspelmeyer, optobook}, leading to a strong dependence of the corresponding variances on the coupling strength (and the intracavity photon number). The impact of the cooling effect gets stronger at higher temperatures. In contrast, at zero temperature there is a very weak dependence of the variances on $g_1$, so that in that case the averages always dominate. Away from the zero-temperature limit, the contribution of the variances to the QFI develops a peak at a particular intra-cavity photon number, reflected as the maxima in the relative error for $g_1$ seen at finite temperatures in Fig. \ref{Fig1}(a).

\begin{figure*}
     \centering
     \begin{subfigure}[b]{0.47\textwidth}
         \centering
         \includegraphics[width=\textwidth]{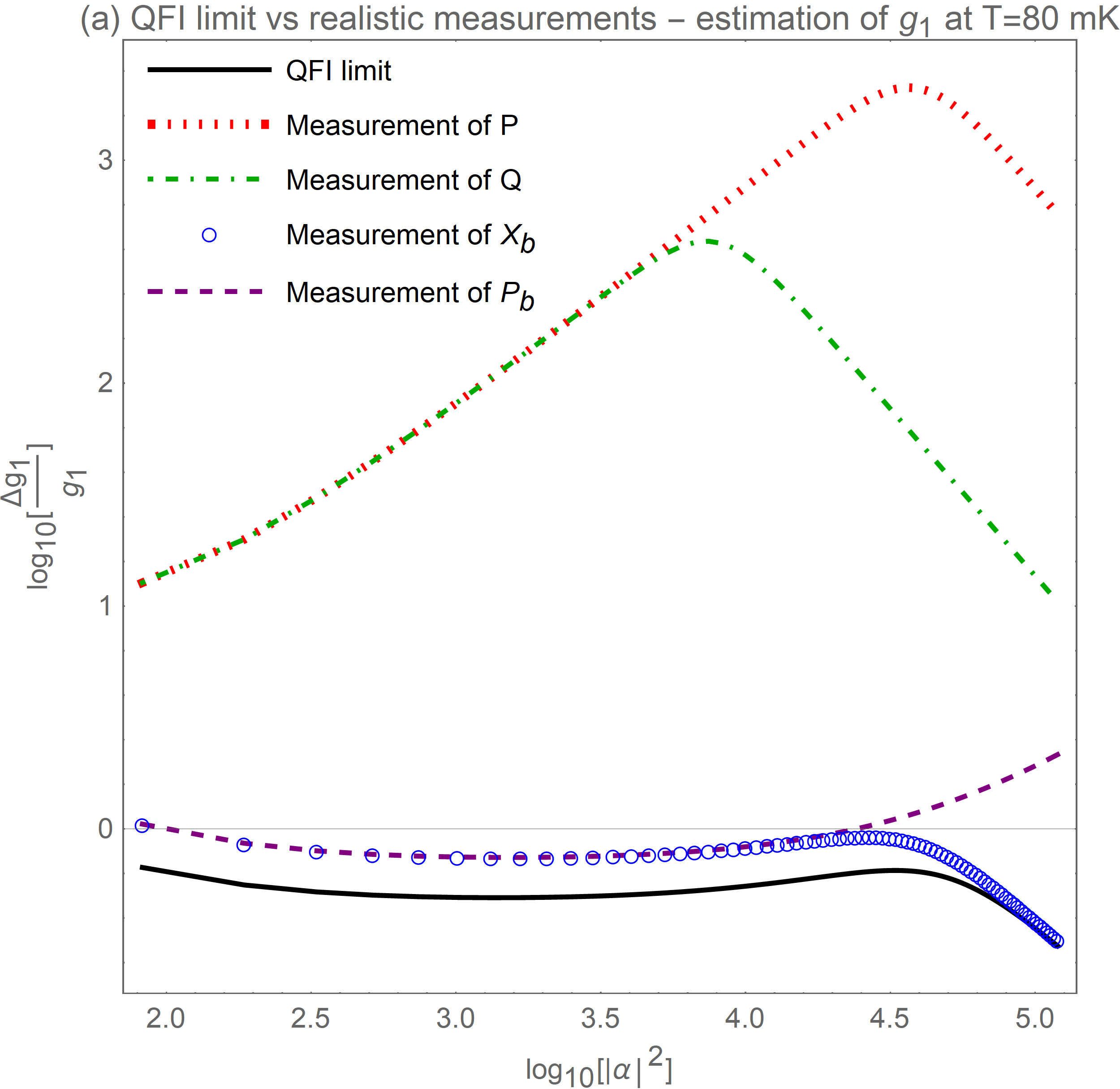}
     \end{subfigure}
     \hfill
     \begin{subfigure}[b]{0.47\textwidth}
         \centering
         \includegraphics[width=\textwidth]{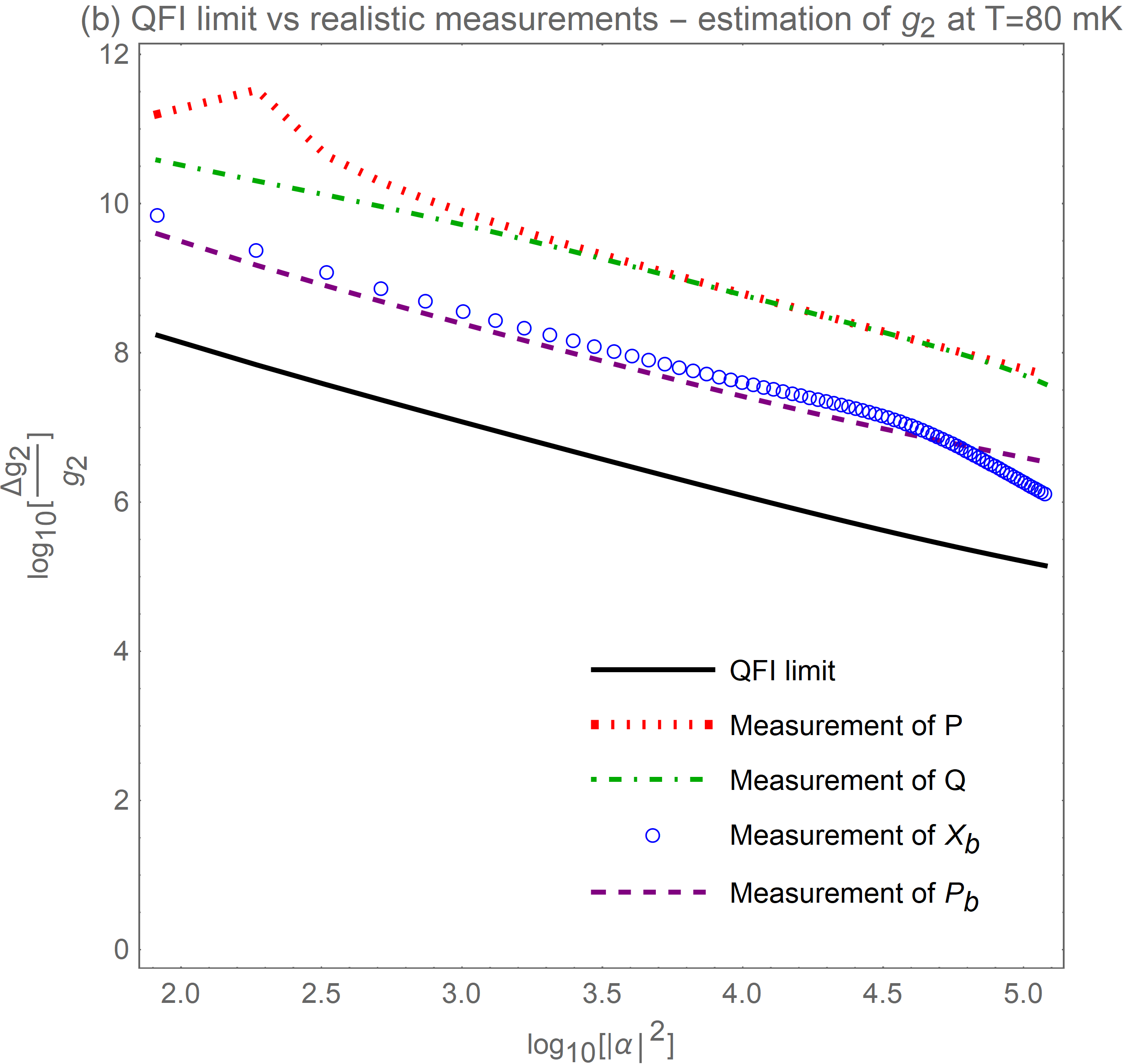}
     \end{subfigure}
     \begin{subfigure}[b]{0.47\textwidth}
         \centering
         \includegraphics[width=\textwidth]{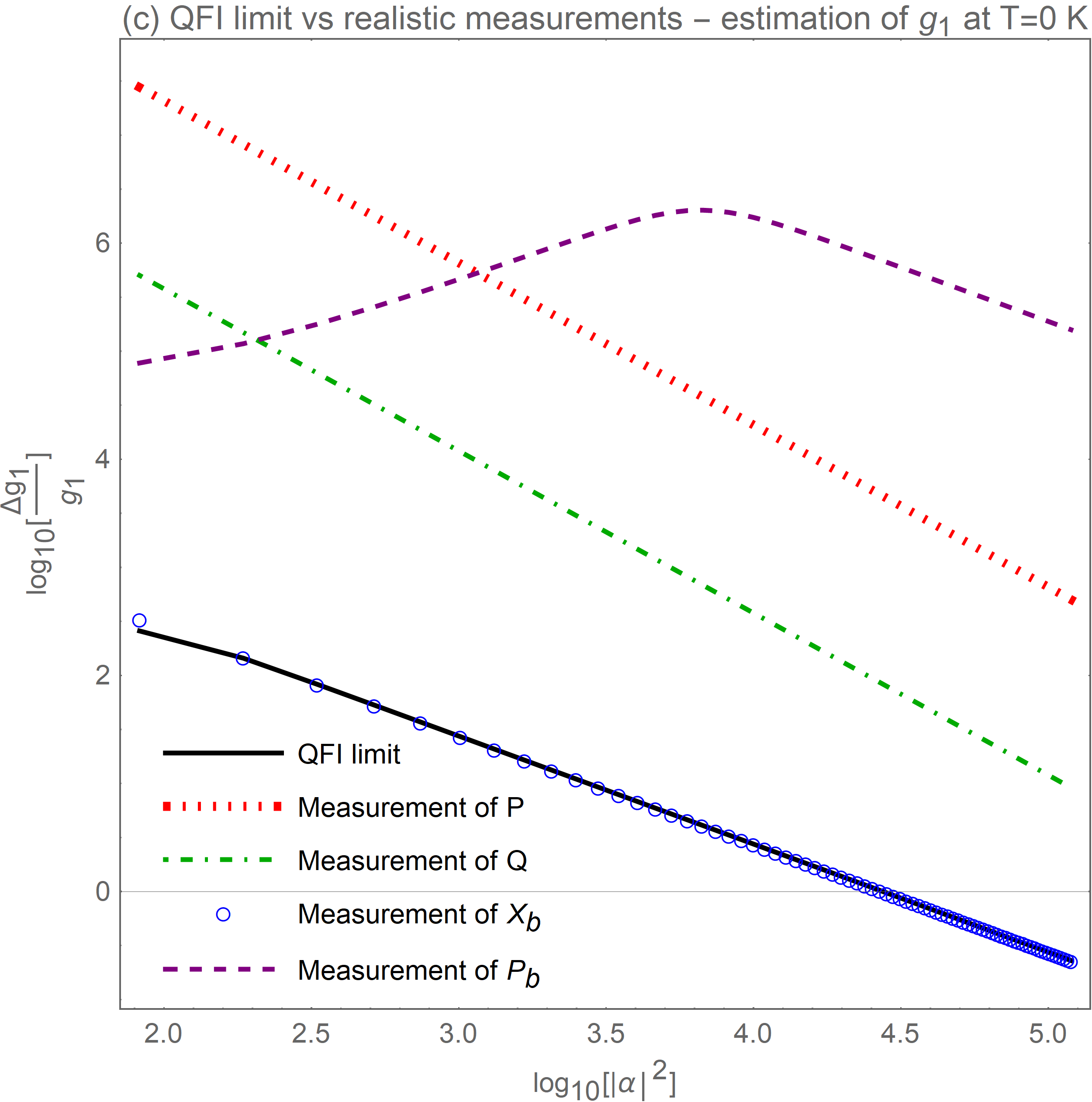}
     \end{subfigure}
     \hfill
     \begin{subfigure}[b]{0.47\textwidth}
         \centering
         \includegraphics[width=\textwidth]{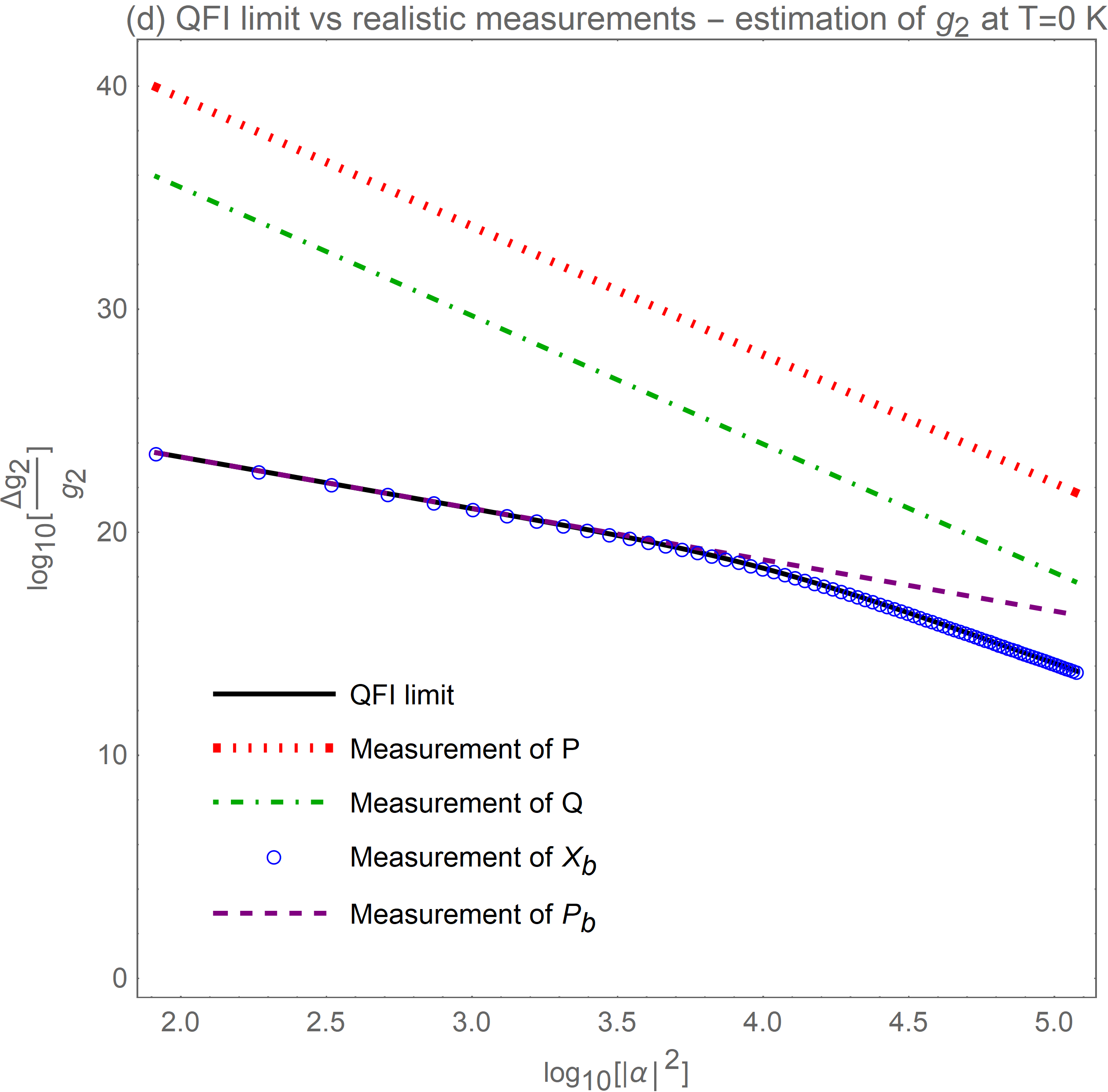}
     \end{subfigure}
      \caption{Log-log plots of relative error bounds for coupling constants against the intracavity photon number, $|\alpha|^2$ as implied by the QFI (black line) and the measurements of $\hat P$ (red dotted line), $\hat Q$ (green dot-dashed line), $\hat X_b$ (blue circles) and $\hat P_b$ (purple dashed line). (a) and (b) are for $g_1$ and $g_2$ in the high temperature scenario respectively, whilst (c) and (d) are for $g_1$ and $g_2$ in the zero-temperature scenario.}
      \label{fig:3}
\end{figure*}

In Fig. \ref{Fig1}(b) we explore the effect of cavity driving and temperature on the estimation precision of the quadratic coupling strength, $g_2$. Overall the relative estimation errors are much higher for $g_2$ than $g_1$, which  is unsurprising as the former is several orders of magnitude smaller.

Interestingly, at all $|\alpha|$ within our allowed range the high temperature scenario predicts the lowest relative errors bound on $g_2$: a hotter mechanical bath gives a better estimation precision for all driving strengths below the instability threshold. This can be traced back to the fact that, also in the estimation of $g_2$, the information content of the variances is again higher than that of the averages at lower driving strengths, and in the high temperature case  it remains so for all driving strengths up until the instability threshold.  The effect gets weaker as the temperature is reduced, and in the $T=0$ case a crossover is seen with the contribution of the averages eventually becoming dominant (see Fig. \ref{fig:5} in Appendix \ref{Appendix}). The overall result is that the relative error bounds on $g_2$ decrease monotonically with increasing drive across the parameter range studied.

In Fig. \ref{Fig2} we compare global and local QFIs for $g_1$ (Fig. \ref{Fig2}(a)) and $g_2$ (Fig. \ref{Fig2}(b)). In a nutshell, we find that the majority of information about the coupling parameters is contained in the reduced state of the mechanics. Note that, in standard optomechanical experiments, measurements are typically performed on the light mode. Nevertheless, our results suggest that significantly more information about the couplings might be available by probing the mechanical motion more directly. 

For $g_1$ the uncertainties found from either just the mechanical subsystem, or just the optical subsystem only drop monotonically with drive strength at $T=0$, matching what happens with the full system. In this case, even at $80$mK the uncertainties obtained from the reduced state of the mechanics almost match those from the full system. In contrast, for  $g_2$ the uncertainty obtained from the state of just the mechanics only reaches that achieved with the full state in the zero temperature limit.

Finally, in Fig. \ref{fig:3} we show how some realistic measurements perform in comparison to the ultimate limits given by the QFI. The figure shows that, out of the measurements considered, the mechanical position almost always does best at estimating the coupling parameters. The ultimate limits to estimation precision of the coupling strengths can only be approached at low and intermediate drive strengths via measurement of $\hat X_b$ at zero temperature. For higher temperatures this limit is approached for $g_1$ at very high intracavity photon numbers, whilst for $g_2$ it is never achieved.
\section{Conclusions}
\label{summary}
We employed local QET to the problem of estimating linear and quadratic coupling parameters in driven-dissipative optomechanics. For experimentally realistic values of the model parameters, inspired by Ref.~\cite{goodcavity}, we have found that it is considerably easier to estimate the linear coupling strength than the quadratic one. Our analysis has also showed that the best strategy for estimating the coupling parameters can be well approximated by a direct measurement of the mechanical position $\hat X_b$.

Exploring the effect of temperature on the estimation precision of the coupling strengths, we found that higher temperatures are not always detrimental to the estimation performance. The effect of temperature is particularly striking when analyzing the estimation of the quadratic coupling parameter: in this case we found that a hotter mechanical bath ($T=80$mK) resulted in a higher estimation precision for all drive strengths below the instability threshold. In contrast, in the case of the linear coupling strength the effect of temperature is most significant at lower driving. Past a certain drive strength, better estimation precision for the linear coupling parameter is instead achieved at lower temperatures. 
\section*{Acknowledgments}
T.T. acknowledges support from the University of
Nottingham via a Nottingham Research Fellowship. K.S. and A.D. acknowledge support from the University of Nottingham. A.D.A. was supported through a Leverhulme Trust Research Project Grant (RPG-2018-213).

\appendix
\section{Covariance matrix language}
\label{covariancematrixlanguage}
In this section we introduce the covariance matrix language. This is particularly convenient for Gaussian states, which can be fully characterised by their first and second moments \cite{serafini}.

Let us consider a system of $N$ bosonic modes, described by a vector of quadratures $\hat{\boldsymbol{R}}=(\hat X_1,\hat P_1,...,\hat X_N,\hat P_N)$ \cite{gaussianstates}. The commutator between any two quadrature operators is given by the corresponding element of a matrix of commutators $\boldsymbol{W}$ which, by construction, satisfies $\boldsymbol{W^T}=-\boldsymbol{W}$. Symbolically, the elements of $\boldsymbol{W}$ are 
\begin{align}
\label{Wij}
    W_{ij}=[\hat R_i,\hat R_j].
\end{align}

In the main text we are dealing with open system dynamics which can be approximated via a bilinear master equation of the general form
\begin{align}
\label{genformme2}
    \dot{\rho}(t)&=-\frac{i}{\hbar}[H',\rho(t)]\nonumber\\
    &+\sum_{ij}\frac{\gamma_{ij}}{2}\left[2\hat R_i\rho(t)\hat R_j-\{\hat R_j\hat R_i,\rho(t)\}\right],
\end{align}
where 
\begin{align}
\label{linhamgen}
    H'=\frac{1}{2}\sum_{ij}H_{ij}\hat R_{i}\hat R_{j}=\frac{1}{2}\hat{\boldsymbol{R}}^T\boldsymbol{H}\hat{\boldsymbol{R}}
\end{align}
is the bilinear Hamiltonian, $\boldsymbol{H}$ is the Hamiltonian matrix and $\gamma_{ij}$ is the corresponding damping rate \cite{serafini}.  Eq. (\ref{linhamgen}) is the general expression for a strictly quadratic Hamiltonian. Without loss of generality we may assume that $\boldsymbol{H}$ is a symmetric matrix, that is $\boldsymbol{H}=\boldsymbol{H}^T$. If the master equation admits a steady state, the latter will be Gaussian provided that the additional condition $\boldsymbol{H}>0$ is satisfied \cite{gaussianstates, serafini}. 

\begin{figure*}[t]
     \centering
         \includegraphics[width=.475\linewidth]{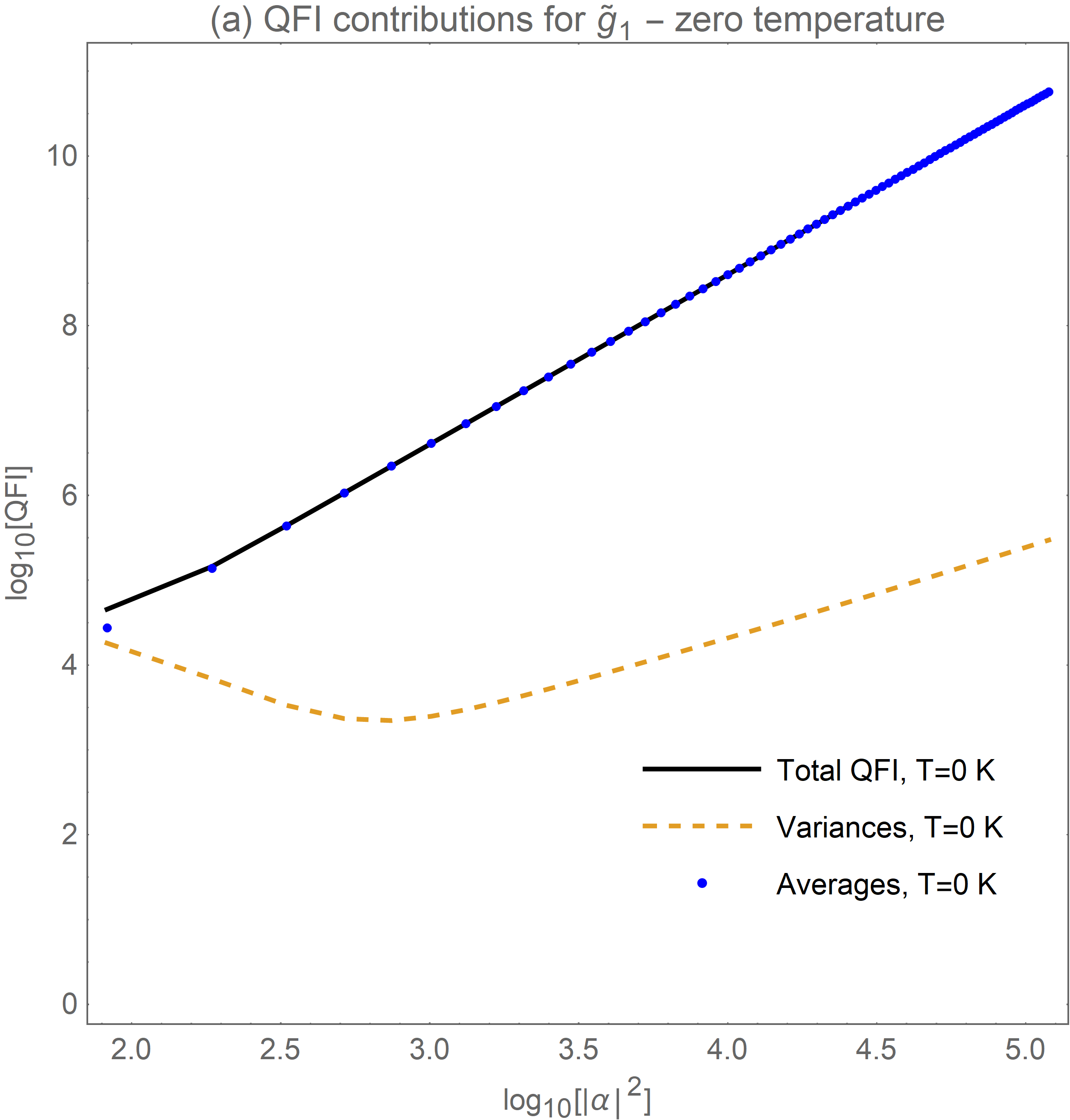}
     \hfill
         \includegraphics[width=.475\linewidth]{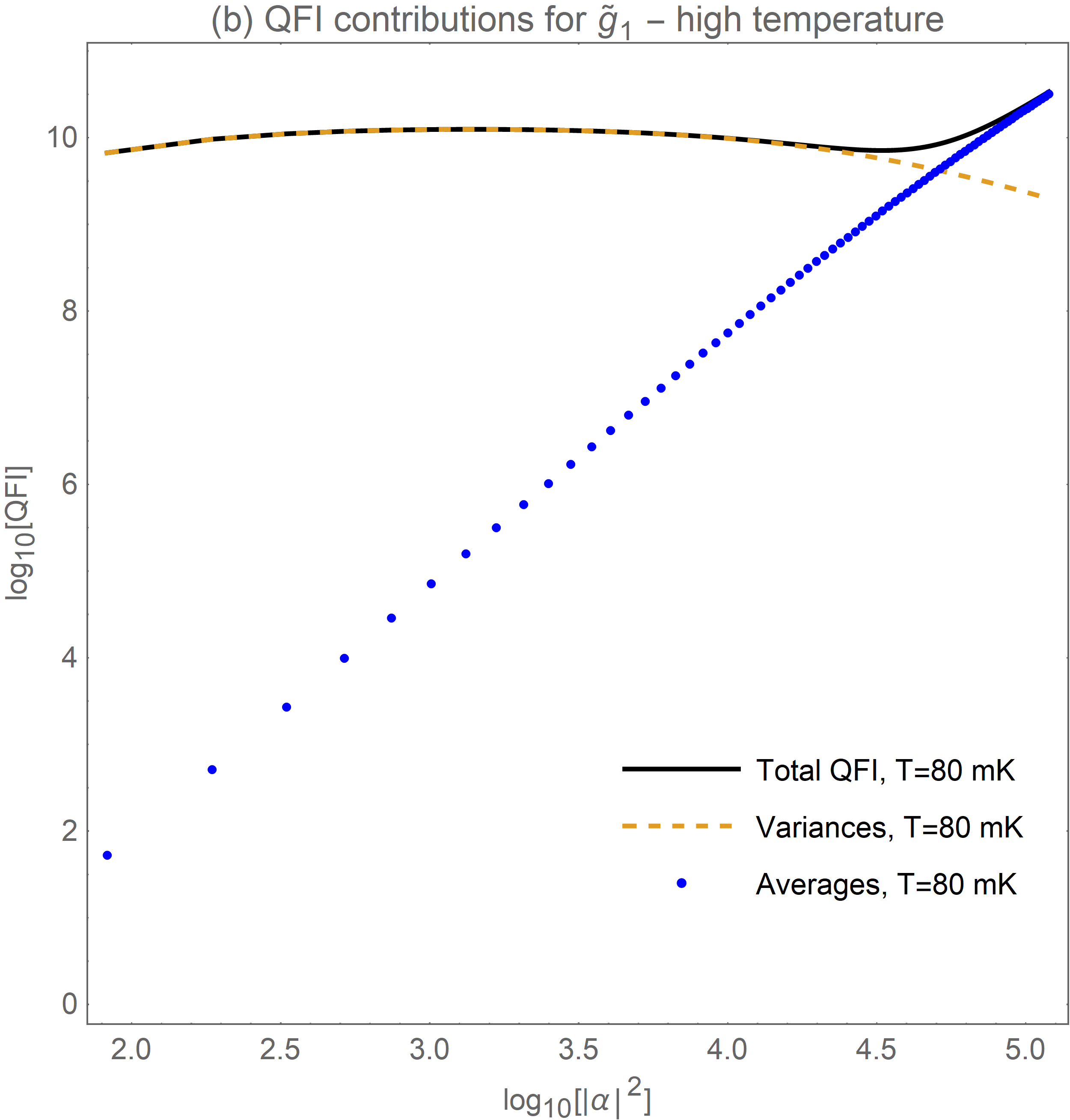}
     \caption{Log-log plots comparing the contributions from the variances (yellow dashed line) and the averages (blue dots) to the QFI (black line) for $g_1$ in (a) the zero temperature scenario and (b)the high temperature scenario.}
     \label{fig:4}
\end{figure*}

\begin{figure*}
     \centering
         \includegraphics[width=.475\linewidth]{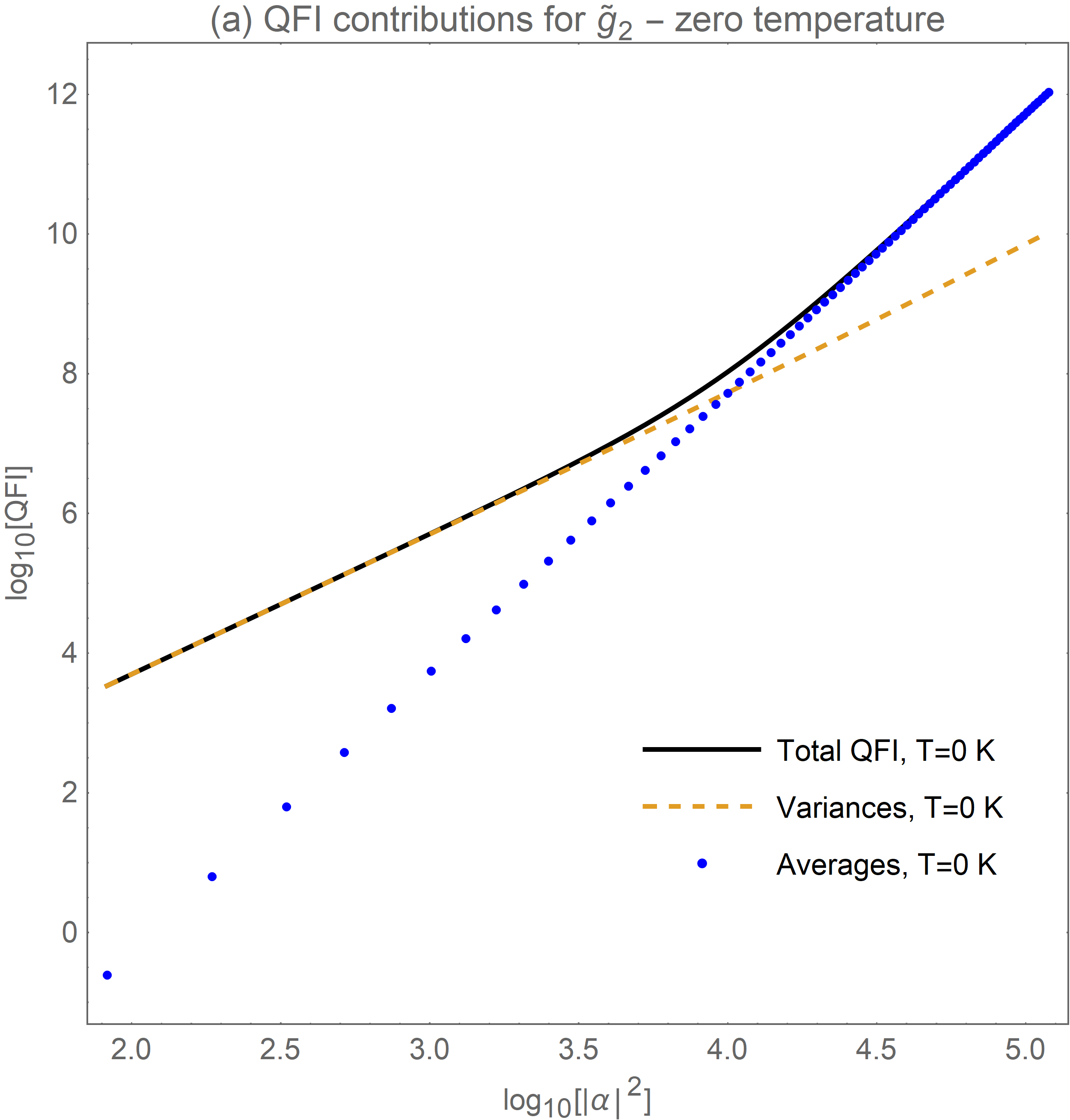}
     \hfill
         \includegraphics[width=.475\linewidth]{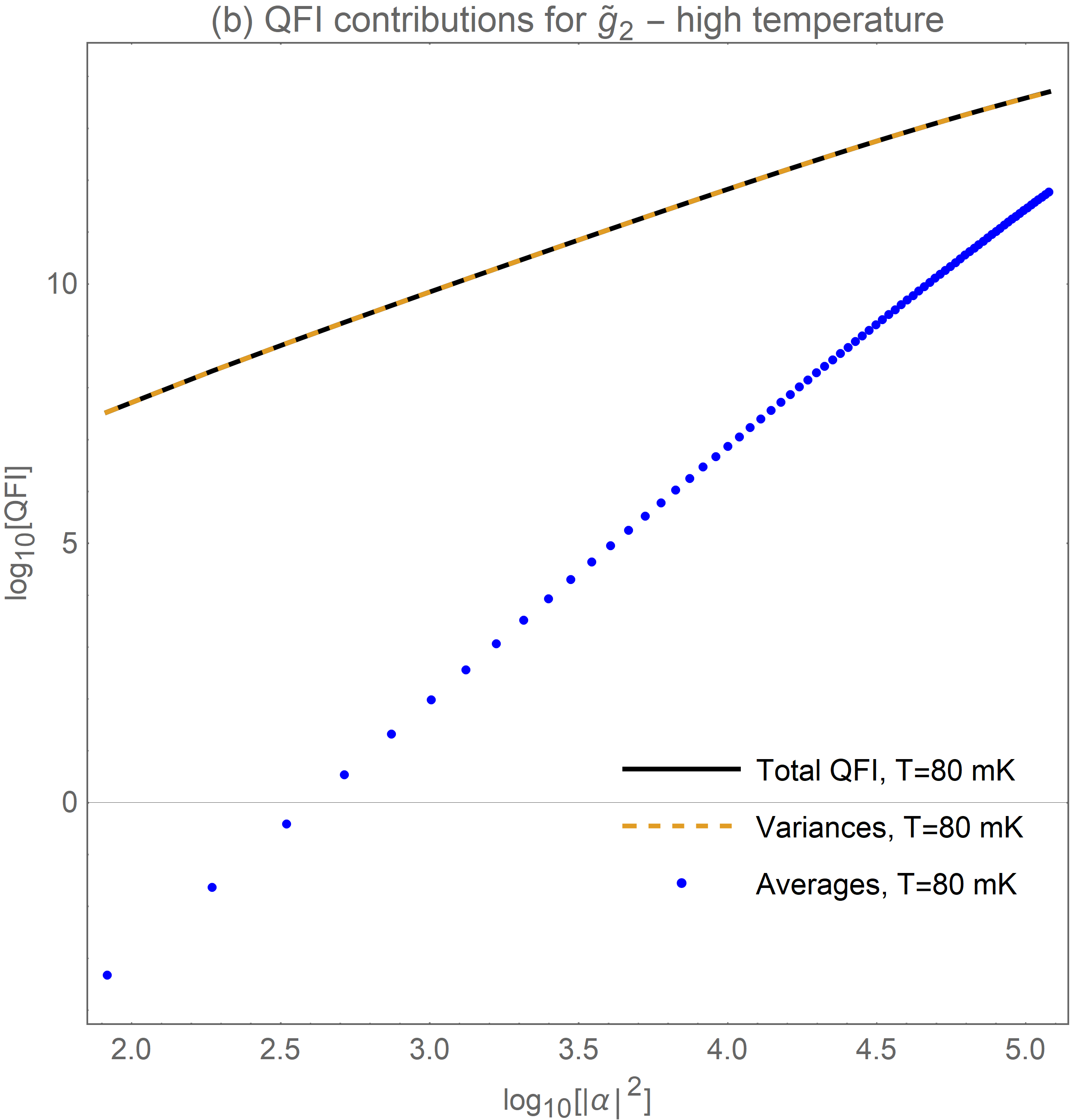}
     \caption{Log-log plots comparing the variances (yellow dashed line) and averages (blue dots) contributions to QFI (black line) for $\tilde{g}_2$ at (a) zero temperature and (b)  in the high temperature scenario.}
     \label{fig:5}
\end{figure*}

The first moments  $\boldsymbol{R}_0$ of a quantum state $\rho(t)$ form a vector of average values defined as \cite{moments,nongaussian}
\begin{align}
 \boldsymbol{R}_0=\langle \hat{\boldsymbol{R}}\rangle=Tr[\rho(t)\hat{\boldsymbol{R}}],
\end{align}
while the second moments $\boldsymbol{\sigma}$ are encoded in the covariance matrix with elements \cite{gaussianstates, serafini, moments}
\begin{align}
\sigma_{kl}=\frac{1}{2}\langle \{\hat R_k,\hat R_l\}\rangle -\langle \hat R_k\rangle \langle \hat R_l\rangle.
\end{align}
Moreover, given a master equation of the form (\ref{genformme2}) the equation of motion for the average value of a generic observable $\hat A$ can be deduced
\begin{align}
    \frac{d}{dt}\langle \hat A\rangle\equiv Tr[\dot{\rho}(t)\hat A]=\frac{i}{\hbar}\langle[H' ,\hat A]\rangle+\langle \mathcal{D}^\dagger(\hat A)\rangle,
\end{align}
where
\begin{align}
    \mathcal{D}^\dagger(\hat A)=\sum_{ij}\frac{\gamma_{ij}}{2}(\hat R_j[\hat A, \hat R_i]+[\hat R_j,\hat A]\hat R_i)
\end{align}
is the dissipator \cite{breuer}. Using the same convention, we find that the vector of first moments evolves according to \cite{gaussianstates, serafini}:
\begin{align}
    \dot{\boldsymbol{R}}_0=-\frac{i}{\hbar}\boldsymbol{WH}\boldsymbol{R}_0+\boldsymbol{W\gamma_A}\boldsymbol{R}_0,
\end{align}
where $\boldsymbol{\gamma}$ is a matrix with elements $\gamma_{ij}$ that has been conveniently decomposed into its symmetric and antisymmetric parts:
\begin{align}
\label{gammaS}
    \boldsymbol{\gamma_S}&=\frac{\boldsymbol{\gamma}+\boldsymbol{\gamma}^T}{2},\\
\label{gammaA}
    \boldsymbol{\gamma_A}&=\frac{\boldsymbol{\gamma}-\boldsymbol{\gamma}^T}{2}.
\end{align}
Instead, the covariance matrix obeys the following equation of motion
\begin{align}
    \dot{\boldsymbol{\sigma}}&=(-\frac{i}{\hbar}\boldsymbol{WH}+\boldsymbol{W\gamma_A})\boldsymbol{\sigma}+\boldsymbol{\sigma}(\frac{i}{\hbar}\boldsymbol{HW}+\boldsymbol{\gamma_AW})\nonumber\\
    &+\boldsymbol{W\gamma_SW}.
\end{align}
The steady state covariance matrix, $\bar{\boldsymbol{\sigma}}$, can thus be found by solving the Lyapunov equation
\begin{align}
\label{Lyapunovsolve}
    B^T\bar{\boldsymbol{\sigma}}+\bar{\boldsymbol{\sigma}}B=C,
\end{align}
where
\begin{align}
\label{B2}
    B&=\frac{i}{\hbar}\boldsymbol{HW}+\boldsymbol{\gamma_AW},\\
    C&=-\boldsymbol{W\gamma_SW}.
\end{align}
In order for the steady state covariance matrix to describe a physical state $\rho(t)$ it must be a real, symmetric and a positive semi-definite matrix \cite{gaussianstates, serafini}.  The semi-positivity requirement is satisfied provided that $\boldsymbol{B}$ is \textit{stable}, i.e. the real parts of the eigenvalues of $\boldsymbol{B}$ are all negative \cite{gaussianstates}. Additionally, the Robertson-Schr\"odinger uncertainty relation must be satisfied, i.e. 
\begin{align}
\label{Robertson}
    2\bar{\boldsymbol{\sigma}}+\boldsymbol{W}\geq0,
\end{align}
The inequality (\ref{Robertson}) is in fact a necessary and sufficient condition for $\bar{\boldsymbol{\sigma}}$ to represent the steady state covariance matrix of a Gaussian state \cite{gaussianstates, serafini}. In our case, assuming that $\boldsymbol{B}$ is indeed stable, Eq.~(\ref{Robertson}) can be further simplified to $\boldsymbol{W}\boldsymbol{\gamma}^T\boldsymbol{W}\geq0$, which is equivalent to the simpler condition $\boldsymbol{\gamma}\geq0$. It is easy to check that the latter is always satisfied for the choice of $\boldsymbol{\gamma}$ adopted in the main text.
\section{Non-monotonic behavior of the QFI}
\label{Appendix}
In this Appendix we explore the origins of the non-monotonic behavior of the QFI when mechanical temperature is introduced, by investigating how the different contributions to the QFI (those due to the averages and the variances) behave in the zero and high temperature cases. 

In our set-up, the coupling strengths have dimensions of frequency, whilst the averages and the steady state covariance matrix are dimensionless. Therefore, as the QFI is a dimensionful quantity that involves derivatives with respect to these parameters in its definition, it will have dimensions of the inverse frequency squared. We can remove the complication with units by switching to the dimensionless linear and quadratic coupling strengths defined as $\tilde{g}_1=g_1/\omega_m$ and $\tilde{g}_2=g_2/\omega_m$, respectively. It is straightforward to show that the QFIs for the dimensionless and the original coupling strengths are related via $\tilde{I}_{ii}=\omega_m^2 I_{ii}$ for $i=\{1,2\}$. 

Recall that non-monotonic behavior of the QFI was only observed in the high (and low) temperature scenarios in the case of $\tilde{g}_1$. In order to understand this behavior we compare the contributions from the variances and averages (as defined below eqn.\ \eqref{QFIM}) to the QFI in the cases of $\tilde{g}_1$ and $\tilde{g}_2$, separately. 

Figure \ref{fig:4} shows the evolution of the QFI contributions with drive strength for $\tilde{g}_1$. In the zero temperature limit (Fig. \ref{fig:4}(a)) the contribution of the averages dominates at all $|\alpha|$. Thus, here the majority of information about $g_1$ is always encoded in the averages. In contrast, in the high temperature scenario we observe a crossover of the contributions from the variances and the averages at $\log_{10}(|\alpha|^2)\sim 4.7$. The variances encode most of the QFI at lower drive strengths, but the contribution of the averages grows monotonically until it eventually becomes dominant. The contribution of the variances increases with drive before reaching a maximum and then declining (albeit rather slowly).

The important role of the variances at lower drive strengths for the high temperature case can be associated with the cavity back-action cooling effect \cite{Aspelmeyer,optobook} whereby the mechanical variances are progressively reduced below their values in thermal equilibrium as the drive strength is increased. Thus we naturally expect the variances to encode more information about the coupling as the temperature is increased and the impact of the back-action on the mechanical variances becomes more important. Furthermore, the back-action cooling effect saturates at very strong drive strengths \cite{goodcavity,Aspelmeyer} and hence it is perhaps no surprise that the contribution of the variances eventually starts to decline.

In Fig. \ref{fig:5} we investigate the QFI contributions for $\tilde{g}_2$. Here, a crossover is only observed in the zero temperature scenario at $\log_{10}(|\alpha|^2)\sim 4$. However, unlike in the $\tilde{g}_1$ case, this crossover of the variances and averages contributions does not give rise to any non-monotonic behavior. In the high temperature scenario the majority of information about the parameter is contained in the variances over the whole range of $|\alpha|$ studied.

\end{document}